\documentclass[sigconf]{acmart}
\AtBeginDocument{%
  }

\copyrightyear{2025}
\acmYear{2025}
\setcopyright{acmlicensed}\acmConference[MMAsia '25]{ACM Multimedia Asia}{December 9--12, 2025}{Kuala Lumpur, Malaysia}
\acmBooktitle{ACM Multimedia Asia (MMAsia '25), December 9--12, 2025, Kuala Lumpur, Malaysia}
\acmDOI{10.1145/3743093.3771038}
\acmISBN{979-8-4007-2005-5/2025/12}
\usepackage{graphicx}

\usepackage{amsmath,amssymb,amsfonts}
\usepackage{algorithmic}
\usepackage{textcomp}
\usepackage{xcolor}
\usepackage{subcaption}
\usepackage{booktabs}
\usepackage{multirow}

\begin{document}

\title{FabasedVC: Enhancing Voice Conversion with Text Modality Fusion and Phoneme-Level SSL Features}

\author{Wenyu Wang}
\affiliation{%
  \institution{School of Software Engineering, Xi'an Jiaotong University}
  \city{Xi'an}
  \country{China}
}
\affiliation{%
  \institution{SYKI-SPEECH Team}
  \city{Xi'an}
  \country{China}
}
\email{wenyu.wang@stu.xjtu.edu.cn}

\author{Zhetao Hu}
\affiliation{%
  \institution{School of Software Engineering, Xi'an Jiaotong University}
  \city{Xi'an}
  \country{China}
}
\affiliation{%
  \institution{SYKI-SPEECH Team}
  \city{Xi'an}
  \country{China}
}
\email{2219904645@qq.com}

\author{Yiquan Zhou}
\affiliation{%
  \institution{School of Software Engineering, Xi'an Jiaotong University}
  \city{Xi'an}
  \country{China}
}
\affiliation{%
  \institution{SYKI-SPEECH Team}
  \city{Xi'an}
  \country{China}
}
\affiliation{%
  \institution{AI Platform Department, bilibili}
  \city{Shanghai}
  \country{China}
}
\email{yiqian.zhou@stu.xjtu.edu.cn}
\authornote{Corresponding author}

\author{Jiacheng Xu}
\affiliation{%
  \institution{School of Software Engineering, East China Normal University}
  \city{Shanghai}
  \country{China}
}
\email{xujiacheng28@outlook.com}

\author{Zhiyu Wu}
\affiliation{%
  \institution{School of Computer Science and Artificial Intelligence, Fudan University}
  \city{Shanghai}
  \country{China}
}
\email{wuzy24@m.fudan.edu.cn}

\author{Chen Li}
\affiliation{%
  \institution{School of Software Engineering, Xi'an Jiaotong University}
  \city{Xi'an}
  \country{China}
}
\email{cclidd@xjtu.edu.cn}

\author{Shihao Li}
\affiliation{%
  \institution{Division of Music and Audio, Union Wheatland Culture and Media Ltd.}
  \city{Chengdu}
  \country{China}
}
\email{yiseho@yiseho.com}

\renewcommand{\shortauthors}{Wenyu Wang et al.}

\begin{abstract}
In voice conversion (VC), it is crucial to preserve complete semantic information while accurately modeling the target speaker's timbre and prosody. This paper proposes FabasedVC to achieve VC with enhanced similarity in timbre, prosody, and duration to the target speaker, as well as improved content integrity. It is an end-to-end VITS-based VC system that integrates relevant textual modality information, phoneme-level self-supervised learning (SSL) features, and a duration predictor. Specifically, we employ a text feature encoder to encode attributes such as text, phonemes, tones and BERT features. We then process the frame-level SSL features into phoneme-level features using two methods: average pooling and attention mechanism based on each phoneme's duration. Moreover, a duration predictor is incorporated to better align the speech rate and prosody of the target speaker. Experimental results demonstrate that our method outperforms competing systems in terms of naturalness, similarity, and content integrity. We strongly recommend that readers listen to our samples.\footnote{https://fabased-vc.github.io/fabasedvc/}
\end{abstract}



\maketitle

\section{Introduction}
Voice conversion (VC) is a technology that modifies the style and timbre of a source speaker's speech to mimic that of a target speaker, while maintaining the original linguistic content\cite{mohammadi2017overview,liu2020transferring}. Both Voice Conversion and Text-to-Speech (TTS) fall under the umbrella of speech processing technologies; however, they differ in their primary focuses. TTS systems\cite{kim2021conditional,kong2023vits2,du2024cosyvoice} primarily focus on generating natural and fluent speech expressions from text. VC systems aims to retain the content information of the source speech while preserving non-linguistic elements such as emotion and expressiveness. It seeks to generate converted speech that combines these elements with the timbre information of the target speaker.

Based on the aforementioned objectives, the key to achieving successful voice conversion is to disentangle speaker independent information and speaker dependent timbre information from source and target speech. On this basis, the two types of information can be integrated, thereby achieving the effect of voice conversion. However, several challenges remain in the field. Firstly, the speaker-independent features extracted are entirely derived from the source speech, which means that their duration and prosody closely follow those of the source. However, each speaker possesses a unique speaking style. Following the prosody, pauses, and duration of the source speech strictly when generating converted speech does not adequately reflect the relevant characteristics of the target speaker. This can reduce the perceived similarity to the target speaker among listeners. Secondly, existing voice conversion systems encounter challenges in preserving the integrity of the content in the generated speech.  This in turn affects the semantic accuracy and naturalness of the generated speech. Lastly, current disentanglement methods are inadequate in effectively removing timbre information, resulting in issues of timbre leakage during voice conversion tasks.

Text-to-Speech technology can complement and mitigate some of the shortcomings present in VC technologies. By integrating commonly used semantic-related information from TTS through front-end processing methods, such as ASR, can be leveraged to enhance the content integrity of converted speech in VC systems. The duration predictors utilized in TTS can better simulate speech durations that align more closely with the characteristics of the target speaker.

In this paper, we propose a novel end-to-end VC system named FabasedVC, which integrates TTS related technologies to address the three aforementioned shortcomings of current VC technologies. Specifically, this system enhances existing VC frameworks by introducing supplementary textual modality information, processing features at the phoneme level, and integrating a duration predictor. Firstly, the system employs ASR and related annotation tools to add various character and phoneme-level textual modality annotations to the existing VC system, thereby improving content completeness. Secondly, We utilize a Forced Aligner tool to obtain phoneme-level timestamps, which we then utilize to convert the disentangled SSL features from the frame level to the phoneme level. This conversion is achieved through phoneme-level average pooling and attention mechanisms. Compared to frame-level features, phoneme-level features provide the advantage of modifiable duration and significantly enhance the disentanglement of speaker timbre information. Lastly, We incorporate a duration predictor into the model which allows the duration of each phoneme to be re-predicted during inference. The re-prediction uses the encoded phoneme-level features combined with the target speaker’s information. When the re-prediction disabled, VC matches the source speech duration; when enabled, it adjusts the duration to improve similarity. Experimental results demonstrate that FabasedVC effectively preserves the content information from the source speech while achieving a high degree of similarity in timbre, duration, and prosody with the target speaker. The contributions of this paper are as follows:
\begin{itemize}
    \item We have designed a pre-processing program for textual modality annotation, which supplements semantic information during the VC process.
    \item By utilizing pooling and attention mechanisms, we have converted frame-level SSL features to the phoneme-level. This has further separated speaker information and provided the capability to adjust duration in VC.
    \item We have added a duration predictor to the VC system that, when enabled, re-estimates the duration of each phoneme for the target speaker.
\end{itemize}

\section{Related Work}
\subsection{Voice Conversion}
Voice conversion (VC) has witnessed remarkable progress, evolving from traditional statistical models to advanced deep learning approaches. Early VC methods mainly relied on signal processing and statistical techniques, such as Gaussian mixture models (GMMs)~\cite{stylianou1998continuous}, frequency warping~\cite{erro2010voice}, and exemplar-based strategies~\cite{takashima2012exemplar}.These methods focused on transforming the spectral features of the source speaker to match those of the target speaker, but often struggled with flexibility and generalization to unseen data.

With the rapid development of deep learning, VC systems have significantly advanced in terms of modeling capacity and performance. Generative Adversarial Networks (GANs) have been widely adopted for voice conversion, where adversarial training is employed to perform end-to-end voice conversion or enhance the performance of existing frameworks~\cite{li2021starganv2,kaneko2021maskcyclegan}. 
However, due to the instability of GANs, the generated speech lacks clarity. Another effective line of work leverages features derived from Automatic Speech Recognition (ASR) systems, such as phoneme posteriorgrams (PPGs) and bottleneck features (BNFs). These representations help disentangle linguistic content from speaker identity, facilitating more robust and speaker-independent VC~\cite{wang2021enriching,zhao2022disentangling,ning2023expressive}. 
There are also some problems with the intermediate features of ASR: The accuracy and granularity of data annotation affecting the model's performance, and the loss of part of the expressiveness in the features. In recent years, self-supervised learning (SSL) has gained increasing attention due to its strong performance on a wide range of speech processing tasks~\cite{hsu2021hubert,chen2022wavlm}. Researchers have explored various strategies to disentangle the speaker identity from SSL-based representations for VC purposes. For example, adversarial speaker disentanglement~\cite{zhao2023adversarial} introduces an adversarial training framework along with an unannotated external corpus to reduce residual speaker information in SSL features. FreeVC~\cite{li2023freevc} proposes a data augmentation approach based on spectrogram resizing, which distorts speaker-related information during SSL feature extraction to improve model robustness and disentanglement capability. SoftVC~\cite{van2022comparison} further enhances semantic representation learning by predicting distributions over discrete units extracted from SSL models.

\subsection{Prosody Prediction}
Prosody, which encompass rhythm, stress, and intonation, is essential for producing natural and expressive speech. Effective prosody modeling not only improves the fluency and naturalness of synthesized speech, but also enables the expression of various styles and emotions, such as emphasis, questioning, or surprise.

Traditional text-to-speech (TTS) systems often struggled with prosody control due to the reliance on handcrafted features and pipeline architectures. The introduction of end-to-end models like Tacotron~\cite{wang2017tacotron} greatly simplified the synthesis process but exposed new challenges, such as limited controllability over prosody and issues like word skipping or repetition. To mitigate this, text-speech alignment becomes crucial. FastSpeech~\cite{ren2019fastspeech} addresses this by introducing a phoneme duration predictor and a length regulator, allowing explicit control over speaking rate and prosody.However, its dependence on external autoregressive models for alignment supervision constrained its flexibility. To overcome these limitations, models such as Glow-TTS~\cite{kim2020glow} and VITS~\cite{kim2021conditional} adopt Monotonic Alignment Search (MAS), using dynamic programming to learn the alignment between text and speech representations without external models.Furthermore, stochastic duration predictors were introduced during inference to increase prosodic diversity and naturalness. With the rise of large language models (LLMs), many recent approaches adopt discrete codes as intermediate representations for speech synthesis. Models like AudioLM~\cite{borsos2023audiolm} and VALL-E~\cite{wang2023neural} implicitly learn prosody from prompt audio without explicit prosody predictors. In contrast, CosyVoice~\cite{du2024cosyvoice} proposes Supervised Semantic Tokens, encoding prosody, timbre, and style into dedicated token embeddings, which are combined with textual inputs for fine-grained prosody control.

\section{Proposed Methods}
\begin{figure}[htb]
  \centering
  \centerline{\includegraphics[width=8.8cm]{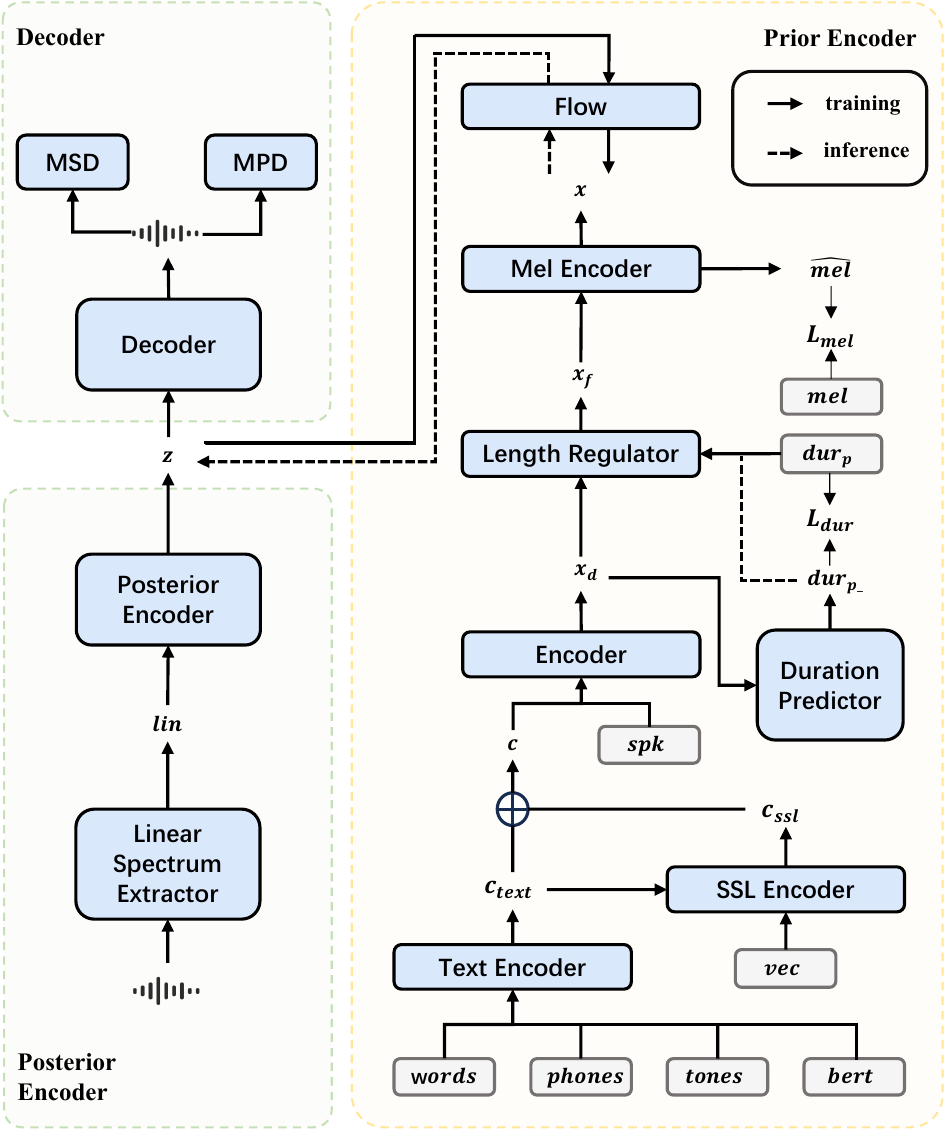}}
\caption{Architecture of FabasedVC.}
\label{fig:res}
\end{figure}

Our proposed end-to-end VC system is shown in Fig.\ref{fig:res}. The backbone of FabasedVC is inspired by VITS\cite{kim2021conditional}, we describe it as comprising a feature extractor and a Conditional Variational Autoencoder (CVAE)\cite{Kingma2014}. The CVAE primarily consisting of a posterior encoder, a prior encoder, and a decoder. The key modifications and contributions of this paper focus on front-end feature processing and the prior encoder. The front-end feature processing supplements a large amount of textual modality information for the speech through a series of automated workflows. In the prior encoder of the VC model, we introduce a text feature encoder, an SSL feature encoder that processes SSL features at the phoneme level, and a duration predictor. First, we will introduce the various features utilized by the model, followed by a detailed description of the specific architecture of the model's encoder and decoder.

\subsection{Features}
\label{ssec:subhead}

\begin{table}
    \caption{Features used for model training and inference.}
    \label{tab:2}
    \renewcommand{\arraystretch}{1.1}
    \begin{tabular}{l@{\hspace{25pt}}l@{\hspace{25pt}}l@{\hspace{25pt}}}
    \hline Feature & Value/example & Type \\
    \hline Words & Chinese characters & Categorical  \\
    Phonemes & sil n i h ao sil & Categorical  \\
    Tones & 0 2 3 0 & Categorical  \\
    WordsBERT & Tensor vector & Numerical \\
    Pframe & 8 9 5 10 33 10 & Numerical \\
    Wframe & 8 14 43 10 & Numerical \\
    W2P & 1 2 2 1 & Numerical \\
    Speaker & speaker1 & Categorical \\
    Contentvec & Tensor vector & Numerical \\
    \hline
    \end{tabular}
\end{table}

In Table \ref{tab:2}, we outline the features utilized for model training and inference, using Mandarin Chinese as a demonstrative example. Words, Phonemes, and Tones are obtained through ASR models and related annotation tools, representing the textual content, phonetic information, and tonal characteristics of speech, respectively. Tonal characteristics refers to the pitch variation patterns of a syllable during pronunciation. This supplementation is based on the fact that Mandarin is a tonal language. For non-tonal languages like English, other semantic features can be used during processing. Some speech voice synthesis systems employ BERT \cite{devlin2018bert} features of text to enhance semantic information\cite{xiao2020improving,zhang2021extracting}. Similarly, we derive WordsBERT features from the Words feature using BERT to further enrich the information.

A forced alignment annotation model can generate timestamps at the phoneme level of speech. By leveraging existing forced alignment annotation models and  applying them to speech data as well as the Words and Phones features, we can extract duration information for each element of Words and Phonemes in the speech. By aligning this duration information with the frames of the spectrogram, we obtain Pframe and Wframe features, which represent the frame lengths corresponding to each character and Phoneme, respectively. The W2P feature indicates the number of phonemes associated with each character in a sentence. The Speaker feature conveys speaker-related information linked to the speech, while the Contentvec\cite{qian2022contentvec} feature refers to the disentangled SSL features extracted from the speech. Similar to other voice conversion systems, we also utilize audio and spectrograms.

The front-end feature processing program we designed can automatically extract all of the aforementioned speech features in a single step during both training and inference.

\subsection{Prior Encoder}
\label{ssec:subhead}

The prior encoder consists of a text feature encoder, an SSL feature encoder, a frame-level prior network, and a duration predictor.

\subsubsection{Text Feature Encoder}

The text encoder comprises four encoding modules, each module encodes Words, Phonemes, Tones, and WordsBERT features separately. These modules produce encodings of the same dimension: $ c_{words} $, $ c_{phones} $, $ c_{tones} $, and $ c_{bert} $. The $ c_{text} $ feature is the sum of these encoding features:

\begin{align}
c_{text} = c_{words} + c_{phones} + c_{tones} + c_{bert} 
\end{align}

\subsubsection{SSL Feature Encoder}

The SSL feature encoder utilizes both phoneme-level average pooling and an attention-based encoding module to reduce the dimensions of the Contentvec feature $ vec $ from the frame level to the phoneme level, resulting in the final representation $ c_{ssl} $.

Phoneme-level average pooling is guided by the Pframe, duration of each phoneme, denoted as $ dur_p $. Given the Contentvec feature matrix $vec \in \mathbb{R}^{n \times f}$, where $n$ denotes the feature dimension and $f$ denotes the frame length, and the phoneme duration vector $dur_p = [d_1, d_2, ..., d_d]$, where $d$ represents the number of phonemes and $\sum_{i=1}^{d} d_i = f$. Based on the number of frames $d_i$ for each phoneme in $dur_p$, average pooling operation is performed on $vec$ to obtain a feature matrix $vec_{dur}$ with dimensions $[n, d]$. The calculation method is as follows:

\begin{align}
vec_{dur}[:, i] = \frac{1}{d_i} \sum_{j=start_i}^{start_i + d_i - 1} vec[:, j],\quad \text{for  } i = 1, ..., d
\end{align}

Where \( start_i = \sum_{k=1}^{i-1} d_k \) represents the starting frame index for the $i$-th phoneme. Finally, by processing $vec_{dur}$ through the encoder, the encoded phoneme-level pooled feature $c_{avg}$ is obtained.

To dynamically encode the $ vec $ features at the phoneme level using text features, we propose an attention-based encoding module. For each phoneme, the output of the text feature encoder, $ c_{text} $, is used as the query $ Q $. The $ vec $ features serve as both the key $ K $ and value $ V $ to integrate the $ vec $ features corresponding to the phoneme frames. Repeats this process for each phoneme to obtain the encoded feature $ c_{att} $. Following the attention mechanism described in \cite{vaswani2017attention}, we employ scaled dot-product operations as the similarity measure.

\begin{figure}[htb]
  \centering
  \centerline{\includegraphics[width=7.8cm]{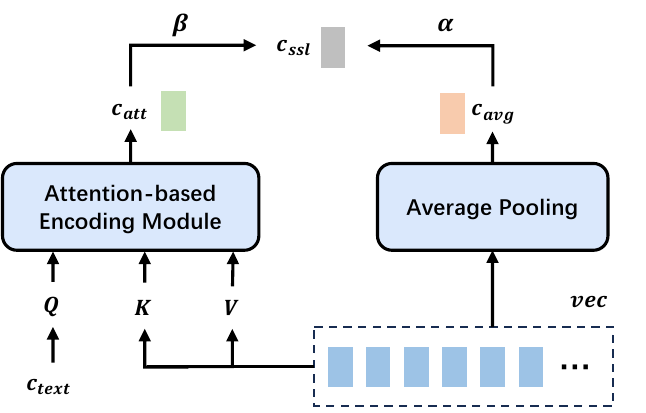}}
\caption{Architecture of SSL Encoder.}
\label{fig:ssl}
\end{figure}
The final SSL encoded feature $ c_{ssl} $ is expressed as:

\begin{align}
c_{ssl} = \alpha c_{avg} + \beta c_{att}
\end{align}

Where $ \alpha $ and $ \beta $ are weighting coefficients. Here, the model performed best with the values set to 1.0 and 0.5, respectively.

Fig.\ref{fig:ssl} shows how the SSL encoder converts all frames of a phoneme in speech into a phoneme-level representation. After this process, the length of the $ c_{ssl} $ feature matches the phoneme feature length of the speech.

\subsubsection{Frame-level Prior Network}

The frame-level prior network expands the phoneme-level features back to the frame-level features $x$ for utilize in subsequent steps. Firstly, we normalizes the sum of $ c_{text} $ and $ c_{ssl} $, resulting in $ c $. By utilizing the speaker information $ spk $ as a condition, it processes $ c $ through an attention-based normalization flow module to produce the encoded phoneme-level feature $ x_d $:

\begin{align}
x_d = encoder(c, spk)
\end{align}

To extend the phoneme-level feature $ x_d $ to the frame level, the phoneme-level feature is replicated according to the frame counts specified by $ dur_{p} $ during training, resulting in the frame-level feature $ x_f $. During inference, to better match the target speaker's duration, the duration predicted by the duration predictor can be utilized to obtain $ x_f $. Details regarding the duration predictor will be discussed in the next subsection.

To enhance the training of the network, we implemented a separate module to predict the mel spectrogram, drawing inspiration from VISinger2 \cite{zhang2022visinger}. The encoder $ x_f $ and $ spk $ serve as inputs to predict the mel spectrogram $ \widehat{mel} $, which is utilized in the loss calculation $ L_{melpre} $ during training. $L_{melpre}$ represents the $L_1$ loss between the predicted mel spectrogram $\widehat{mel}$ and the ground truth mel spectrogram $mel$.
Additionally, $ \widehat{mel} $ is processed through the network and subsequently added back to $ x_f $ to obtain $ x $:

\begin{align}
x = postnet(x_f + prenet(\widehat{mel})) 
\end{align}


\subsubsection{Duration Predictor}

The duration predictor $dp$ re-predicts the logarithm duration $logw_{-}$ using $x_d$ and $spk$. This helps to better match the durations of different target speakers:

\begin{align}
logw_{-} = dp(x, spk)
\end{align}

The duration prediction loss $L_{dur}$ is expressed as the $L_2$ loss between $logw_{-}$ and $logw$. Here, $logw$ represents the actual logarithm duration. 

\subsection{Posterior Encoder}

The posterior encoder is designed to extract the latent representation $ z $ from the linear spectrogram. In line with VITS, we employ non-causal WaveNet\cite{oord2016wavenet} residual blocks as the primary architecture of the posterior encoder.

\subsection{Decoder}

The decoder generates audio waveforms based on the intermediate representation $z$ obtained from the encoder. In our decoder, we utilize the HiFi-GAN architecture \cite{kong2020hifi}, which comprises a generator and two discriminators. The generator accepts $z$ as input and progressively upsamples the data until the generated sequence matches the temporal resolution of the original waveform. The two discriminators are the multi-period discriminator (MPD) and the multi-scale discriminator (MSD), which assess the audio across different periodic windows and scales, respectively. GAN-based training is employed to improve the quality of the reconstructed speech.

\subsection{Final Loss}

The ultimate loss of our proposed model, through CVAE and adversarial training, can be articulated as follows:
\begin{align}
L(G) = L_{rec} + L_{premel} + L_{kl} + L_{dur} + L_{G}
\end{align}
\begin{align}
L(D) = L_{D} 
\end{align}
Where $ L_{rec} $ is the mel spectrogram reconstruction loss, $ L_{kl} $ is the KL divergence loss between the prior and posterior distributions, $ L_{G} $ and $ L_{D} $ are the adversarial training losses associated with the generator and the discriminator.

\section{Experiments}
\subsection{Experimental setup}
\label{ssec:subhead}
\begin{table*}
  \caption{The subjective evaluation results are presented in terms of MOS with 95\% confidence intervals, corresponding to the scenarios of "aishell-to-aishell" and "aishell-to-baker". The objective evaluation results included CER, PER and Cos.Sim.}
  \label{tab:1}
  \renewcommand{\arraystretch}{1.0}
  \begin{tabular}
   {l@{\hspace{15pt}}|c@{\hspace{20pt}}c@{\hspace{20pt}}|c@{\hspace{20pt}}c@{\hspace{20pt}}|c@{\hspace{15pt}}c@{\hspace{15pt}}c@{\hspace{15pt}}}
    \hline & \multicolumn{2}{c|}{ aishell-to-aishell } & \multicolumn{2}{c|}{ databaker-to-aishell } & \multicolumn{3}{c}{ objective } \\
    \midrule
    Approach & Naturalness $\uparrow$ & Similarity
 $\uparrow$ & Naturalness $\uparrow$ & Similarity
 $\uparrow$ & CER $\downarrow$ & PER $\downarrow$ & Cos.Sim
 $\uparrow$ \\
    \midrule
     VQMIVC & $2.21 \pm 0.06$ & $2.55 \pm 0.06$ & $2.26 \pm 0.09$ & $2.61 \pm 0.07$ & $26.64\%$ & $13.61\%$ & $0.6498$ \\
     PPGVC & $3.02 \pm 0.07$ & $2.97 \pm 0.07$ & $2.82 \pm 0.09$ & $2.88 \pm 0.11$ & $12.08\%$ & $5.38\%$ & $0.7020$ \\
     FreeVC & $3.74 \pm 0.08$ & $3.41 \pm 0.07$ & $3.72 \pm 0.09$ & $3.30 \pm 0.11$ & $11.97\%$ & $4.51\%$ &  $0.7349$ \\
     SOVITS-VC & $4.09 \pm 0.06$ & $3.64 \pm 0.07$ & $3.65 \pm 0.11$ & $3.24 \pm 0.08$ & $7.51\%$ & $2.26\%$ & $0.7432$ \\
     Cosyvoice-VC & $3.82 \pm 0.08$ & $4.05 \pm 0.08$ & $3.78 \pm 0.07$ & $3.81 \pm 0.09$ & $10.80\%$ & $4.05\%$ & $0.7836$ \\
     SeedVC & $4.08 \pm 0.06$ & $4.11 \pm 0.08$ & $\mathbf{3.90} \pm \mathbf{0.08}$ & $3.88 \pm 0.07$ & $8.15\%$ & $3.17\%$ & $0.7827$ \\
     FabasedVC& $\mathbf{4.43} \pm \mathbf{0.06}$ & $\mathbf{4.49} \pm \mathbf{0.05}$ & $3.85 \pm 0.08$ & $\mathbf{4.27} \pm \mathbf{0.07}$ & $\mathbf{6.27\%}$ & $\mathbf{2.22\%}$ & $\mathbf{0.8088}$ \\
     \midrule
      - SSL Pooling & $3.81 \pm 0.06$ & $4.29 \pm 0.06$ & $3.53 \pm 0.05$ & $4.09 \pm 0.08$ & $19.19\%$ & $8.48\%$ & $0.7925$ \\
      - SSL Attention & $3.96 \pm 0.06$ & $4.34 \pm 0.06$ & $3.62 \pm 0.06$ & $4.16 \pm 0.07$ & $9.45\%$ & $3.22\%$ & $0.7994$ \\
      - Duration predictor & $4.23 \pm 0.07$ & $4.25 \pm 0.05$ & $3.65 \pm 0.07$ & $4.04 \pm 0.08$ & $11.35\%$ & $4.05\%$ & $0.8078$ \\
    \midrule
  \end{tabular}
\end{table*}

In the experiments, open-source Mandarin data Aishell3 \cite{shi2021aishell} are used to train the model. AISHELL3 contains approximately 85 hours of speech data from around 218 native Mandarin speakers. For evaluation, we randomly selected 10 sentences from each speaker for validation and an additional 10 sentences for testing, using the remaining data for training. Furthermore, we incorporated a dataset featuring a single female speaker from DataBaker\footnote{https://www.data-baker.com/\#/data/index/source} for testing, with the aim of demonstrating the model's robustness in converting speech across different styles.

The annotation information utilized during training and inference is derived from an ASR model \cite{gao2022paraformer} and an internal forced alignment tool. This forced alignment tool employs the text annotations and phonetic information obtained from the ASR model to generate the corresponding timestamps for the text. We utilized pre-trained 192-dimensional Mandarin BERT\cite{cui2021pre} and 256-dimensional ContentVec\cite{qian2022contentvec} models to extract the corresponding features. Through this preprocessing step, we are able to obtain all the features required for training and inference from the speech.

All audio samples were resampled to 44,100 Hz. We employed the Short-Time Fourier Transform (STFT) to compute linear spectrograms and 80-band Mel spectrograms. The dimensions for textual information and SSL-encoded features were set to 192. Our model was trained for 500,000 steps on an A100 GPU with a batch size of 16.

To evaluate the performance of the proposed FabasedVC across various aspects of voice conversion, we compared it with several VC systems. The comparison systems include VQMIVC \cite{wang2021vqmivc}, PPGVC \cite{Liu2021}, FreeVC \cite{li2023freevc}, SOVITS\footnote{https://github.com/svc-develop-team/so-vits-svc}, Cosyvoice\cite{du2024cosyvoice} in voice conversion task, and SeedVC\cite{liu2024zero}. VQMIVC achieves effective VC by disentangling speech representations; PPGVC is a VC system based on the BNF framework; FreeVC employs the VITS framework to provide high-quality waveform reconstruction for VC; and SOVITS is a well-regarded open-source SVC and VC project. Cosyvoice is a highly effective zero-shot TTS model that significantly enhances content consistency and speaker similarity. It also supports voice conversion capabilities. SeedVC achieves high-fidelity, low-leakage zero-shot voice conversion via external timbre perturbation and a diffusion Transformer. For CosyVoice and SeedVC, we used the official open-source models for voice conversion, and we trained other VC comparison models using the same dataset as FabasedVC. Additionally, we conducted ablation studies on the SSL encoder and duration predictor to verify the effectiveness of each component.

\subsection{Evaluation Metrics}
\label{ssec:subhead}

In both subjective and objective evaluations, we assessed naturalness, intelligibility, and speaker similarity. For the subjective evaluation, 20 participants rated the naturalness and similarity of the audio using a 5-point Mean Opinion Score (MOS). We randomly selected four target speakers (two males and two females) from the AISHELL3 dataset and evaluated the system under two scenarios: converting the speech of AISHELL speakers to that of target speakers, and converting the speech of a DataBaker speaker to that of target speakers. The DataBaker dataset features standard pronunciation and stable prosody, whereas speakers in the AISHELL3 dataset exhibit various accents and speaking habits. Evaluating under these two scenarios aims to test the robustness of the system and the prosodic diversity generated for different target speakers. This approach better demonstrates the model's capabilities in modeling duration and prosody. For the objective evaluation, we adopted three metrics: Character Error Rate (CER), Phoneme Error Rate (PER), and Cosine Similarity (Cos.sim). CER was obtained via an ASR model\footnote{https://github.com/wenet-e2e/wenet} to measure the character error rate between the source speech and the converted speech. Similarly, PER measures the phoneme error rate following ASR. Cos.sim is derived from the speaker embeddings of the converted speech, objectively quantifying the similarity between the converted speech and the target speaker. In this paper, we utilized a commonly employed speaker embedding extraction model from PPGVC\footnote{https://github.com/liusongxiang/ppg-vc/blob/main/speaker\_encoder/ckpt} to obtain the speaker embeddings used for testing. Additionally, we evaluated the actual effectiveness of the predicted durations.

\subsection{Results and Analysis}
The results in Table \ref{tab:1} indicate that our model achieves the highest scores on the majority of metrics for both speech naturalness and speaker similarity, surpassing all baseline models, with the exception of a single metric where it ranks second. Even when converting Databaker data to AISHELL speakers, our model maintains a high level of similarity. Despite the significant differences in style and Mandarin proficiency between the source and target speakers, the model achieves it. This clearly demonstrates the role of phoneme-level SSL features and the duration predictor in enhancing speaker similarity. The higher Cos.sim scores further confirm that our model maintains high speaker similarity. Additionally, more comprehensive feature information can facilitate the model in generating speech with higher naturalness. Our proposed model achieves lower CER and PER compared to all other models. This also demonstrates the effectiveness of FabasedVC in addressing the loss of content completeness and better preserving the linguistic content of the source speech. In the ablation experiments, we separately removed the SSL feature encoder modules $c_{att}$ and $c_{avg}$ as well as the duration predictor. It was observed that removing $c_{att}$ or $c_{avg}$ leads to a noticeable decline in both speech naturalness and similarity. This removal also results in some loss of content completeness. Additionally, disabling the duration predictor results in generating speech with the same timing as the source, which can lead to a decrease in perceived speaker similarity.

Table \ref{table:comapre baseline modified} presents a comprehensive analysis of the Relative Duration Deviation (RDD) in the voice conversion task. The RDD metric is computed as:
\begin{align}
\text{RDD} = \frac{|D_{\text{conv}} - D_{\text{ref}}|}{D_{\text{ref}}} \times 100\%
\end{align}
where $D_{\text{conv}}$ and $D_{\text{ref}}$ represent the average phoneme duration of the converted speech and the reference speech, respectively. This formula quantifies the percentage deviation in phoneme duration between the converted utterance and the reference, providing a normalized measure that enables meaningful comparison across speakers with different speaking rates.

As shown, the converted speech exhibits a low average RDD relative to the source speaker, indicating that the model preserves the original temporal structure and speaking rate of the source utterance. At the same time, the average RDD relative to the target speaker is 11.34\%, which is significantly lower than the average RDD\textsubscript{source→target} of 20.85\%. This demonstrates that the model successfully shifts the duration characteristics toward those of the target speaker, achieving a partial but meaningful adaptation without fully sacrificing the source timing. This confirms that the proposed approach successfully leverages the source speaking rate while producing duration characteristics similar to those of the target.
\begin{table}
\caption{Analysis of Relative Duration Deviation (RDD) for voice conversion}
\label{table:comapre baseline modified}
\renewcommand{\arraystretch}{1.0}
\centering
\resizebox{\linewidth}{!}{
\begin{tabular}{l@{\hspace{20pt}}c@{\hspace{20pt}}c@{\hspace{20pt}}c}
\midrule
\textbf{Speakers} & \textbf{RDD\textsubscript{source}} & \textbf{RDD\textsubscript{target}} & \textbf{RDD\textsubscript{source→target}} \\
\midrule
SSB0338 & 0.28\% & 7.32\% & 13.92\% \\
SSB0817 & 4.67\% & 2.64\% & 4.27\% \\
SSB1585 & 6.38\% & 21.87\% & 41.91\% \\
SSB1935 & 1.66\% & 14.52\% & 23.29\% \\
\midrule
Average & 3.25\% & 11.34\% & 20.85\% \\
\midrule
\end{tabular}
}
\end{table}

\section{Conclusion}

In this paper, we propose FabasedVC, a VC system with high timbre similarity and content integrity. We adopted a VITS-based framework for waveform reconstruction. To improve content completeness in voice conversion, we enhanced the prior encoder by introducing textual information through front-end feature processing. Moreover, we processed SSL features from the frame-level to the phoneme-level to disentangle timbre information. We also incorporated a duration predictor to adjust duration. Experimental results demonstrate the high naturalness and similarity achieved by our method. In the future, we will continue to focus on enhancing prosody and content integrity in voice conversion. 

\begin{acks}
This work was supported by the Key Research and Development Program of Shaanxi Province, China, under Grant No. 2024GX-YBXM-556.
\end{acks}

%
%
%
%
\bibliographystyle{ACM-Reference-Format}
\bibliography{sample-base}

@String{Computing = "Computing" }

@ArtifactSoftware{R,
    title = {R: A Language and Environment for Statistical Computing},
    author = {{R Core Team}},
    organization = {R Foundation for Statistical Computing},
    address = {Vienna, Austria},
    year = {2019},
    url = {https://www.R-project.org/},
}

@inproceedings{liu2020transferring,
  title={Transferring Source Style in Nonparallel Voice Conversion},
  author={Liu, Songxiang and Cao, Yuewen and Kang, Shiyin and Hu, Na and Liu, Xunying and Su, Dan and Yu, Dong and Meng, Helen},
  booktitle={International Speech Communication Association (Interspeech)},
  year={2020},
  pages={4721--4725}
}

@article{mohammadi2017overview,
  title={An overview of voice conversion systems},
  author={Mohammadi, S. H. and Kain, A.},
  journal={Speech Communication},
  volume={88},
  pages={65--82},
  year={2017}
}

@article{stylianou1998continuous,
  title={Continuous probabilistic transform for voice conversion},
  author={Stylianou, Yannis and Capp{\'e}, Olivier and Moulines, Eric},
  journal={IEEE Transactions on speech and audio processing},
  volume={6},
  number={2},
  pages={131--142},
  year={1998},
  publisher={IEEE}
}

@article{erro2010voice,
  title={Voice Conversion Based on Weighted Frequency Warping},
  author={Erro, Daniel and Moreno, Antonio and Bonafonte, Antonio},
  journal={IEEE Transactions on Audio, Speech, and Language Processing},
  volume={18},
  number={5},
  pages={922--931},
  year={2010}
}

@inproceedings{takashima2012exemplar,
  title={Exemplar-based Voice Conversion in Noisy Environment},
  author={Takashima, Ryoichi and Takiguchi, Tetsuya and Ariki, Yasuo},
  booktitle={Spoken Language Technology Workshop (SLT)},
  pages={313--317},
  year={2012}
}

@inproceedings{kaneko2021maskcyclegan,
  title={Maskcyclegan-vc: Learning non-parallel voice conversion with filling in frames},
  author={Kaneko, Takuhiro and Kameoka, Hirokazu and Tanaka, Kou and Hojo, Nobukatsu},
  booktitle={ICASSP 2021-2021 IEEE International Conference on Acoustics, Speech and Signal Processing (ICASSP)},
  pages={5919--5923},
  year={2021},
  organization={IEEE}
}

@inproceedings{li2021starganv2,
  title={StarGANv2-VC: A Diverse, Unsupervised, Non-Parallel Framework for Natural-Sounding Voice Conversion},
  author={Li, Y. A. and Zare, A. and Mesgarani, N.},
  booktitle={International Speech Communication Association (Interspeech)},
  year={2021},
  pages={1349--1353}
}

@inproceedings{zhao2022disentangling,
  title={Disentangling content and fine-grained prosody information via hybrid ASR bottleneck features for voice conversion},
  author={Zhao, Xintao and Liu, Feng and Song, Changhe and Wu, Zhiyong and Kang, Shiyin and Tuo, Deyi and Meng, Helen},
  booktitle={ICASSP 2022-2022 IEEE International Conference on Acoustics, Speech and Signal Processing (ICASSP)},
  year={2022},
  pages={7022--7026},
  organization={IEEE}
}

@inproceedings{wang2021enriching,
  title={Enriching source style transfer in recognition-synthesis based non-parallel voice conversion},
  author={Wang, Zhichao and Zhou, Xinyong and Yang, Fengyu and Li, Tao and Du, Hongqiang and Xie, Lei and Gan, Wendong and Chen, Haitao and Li, Hai},
  booktitle={International Speech Communication Association (Interspeech)},
  year={2021}
}

@inproceedings{ning2023expressive,
  title={Expressive-vc: Highly expressive voice conversion with attention fusion of bottleneck and perturbation features},
  author={Ning, Ziqian and Xie, Qicong and Zhu, Pengcheng and Wang, Zhichao and Xue, Liumeng and Yao, Jixun and Xie, Lei and Bi, Mengxiao},
  booktitle={ICASSP 2023-2023 IEEE International Conference on Acoustics, Speech and Signal Processing (ICASSP)},
  pages={1--5},
  year={2023},
  organization={IEEE}
}

@article{hsu2021hubert,
  title={HuBERT: Self-Supervised Speech Representation Learning by Masked Prediction of Hidden Units},
  author={Hsu, Wei-Ning and Bolte, Benjamin and Tsai, Yao-Hung Hubert and Lakhotia, Kushal and Salakhutdinov, Ruslan and Mohamed, Abdelrahman},
  journal={IEEE/ACM Transactions on Audio, Speech, and Language Processing},
  volume={29},
  pages={3451--3460},
  year={2021},
  publisher={IEEE}
}

@article{chen2022wavlm,
  title={Wavlm: Large-scale self-supervised pre-training for full stack speech processing},
  author={Chen, Sanyuan and Wang, Chengyi and Chen, Zhengyang and Wu, Yu and Liu, Shujie and Chen, Zhuo and Li, Jinyu and Kanda, Naoyuki and Yoshioka, Takuya and Xiao, Xiong and others},
  journal={IEEE Journal of Selected Topics in Signal Processing},
  volume={16},
  number={6},
  pages={1505--1518},
  year={2022},
  publisher={IEEE}
}

@inproceedings{li2023freevc,
  title={Freevc: Towards High-Quality Text-Free One-Shot Voice Conversion},
  author={Li, Jingyi and Tu, Weiping and Xiao, Li},
  booktitle={ICASSP 2023-2023 IEEE International Conference on Acoustics, Speech and Signal Processing (ICASSP)},
  pages={1--5},
  year={2023},
  organization={IEEE}
}

@inproceedings{van2022comparison,
  title={A comparison of discrete and soft speech units for improved voice conversion},
  author={van Niekerk, Benjamin and Carbonneau, Marc-Andr{\'e} and Za{\"\i}di, Julian and Baas, Matthew and Seut{\'e}, Hugo and Kamper, Herman},
  booktitle={ICASSP 2022-2022 IEEE International Conference on Acoustics, Speech and Signal Processing (ICASSP)},
  pages={6562--6566},
  year={2022},
  organization={IEEE}
}

@inproceedings{zhao2023adversarial,
  title={Adversarial Speaker Disentanglement Using Unannotated External Data for Self-supervised Representation-based Voice Conversion},
  author={Zhao, Xintao and Wang, Shuai and Chao, Yang and Wu, Zhiyong and Meng, Helen},
  booktitle={2023 IEEE International Conference on Multimedia and Expo (ICME)},
  pages={1691--1696},
  year={2023},
  organization={IEEE}
}

@inproceedings{qian2022contentvec,
  title={Contentvec: An improved self-supervised speech representation by disentangling speakers},
  author={Qian, Kaizhi and Zhang, Yang and Gao, Heting and Ni, Junrui and Lai, Cheng-I and Cox, David and Hasegawa-Johnson, Mark and Chang, Shiyu},
  booktitle={International Conference on Machine Learning},
  pages={18003--18017},
  year={2022},
  organization={PMLR}
}

@inproceedings{kim2021conditional,
  title={Conditional variational autoencoder with adversarial learning for end-to-end text-to-speech},
  author={Kim, Jaehyeon and Kong, Jungil and Son, Juhee},
  booktitle={International Conference on Machine Learning},
  pages={5530--5540},
  year={2021},
  organization={PMLR}
}

@inproceedings{Kingma2014,
  author = {Kingma, Diederik P. and Welling, Max},
  title = {Auto-encoding variational bayes},
  booktitle = {International Conference on Learning Representations},
  year = {2014},
}

@article{vaswani2017attention,
  title={Attention is all you need},
  author={Vaswani, A},
  journal={Advances in Neural Information Processing Systems},
  year={2017}
}

@article{zhang2022visinger,
  title={Visinger 2: High-fidelity end-to-end singing voice synthesis enhanced by digital signal processing synthesizer},
  author={Zhang, Yongmao and Xue, Heyang and Li, Hanzhao and Xie, Lei and Guo, Tingwei and Zhang, Ruixiong and Gong, Caixia},
  journal={arXiv preprint arXiv:2211.02903},
  year={2022}
}

@article{oord2016wavenet,
  title={Wavenet: A generative model for raw audio},
  author={Oord, Aaron van den and Dieleman, Sander and Zen, Heiga and Simonyan, Karen and Vinyals, Oriol and Graves, Alex and Kalchbrenner, Nal and Senior, Andrew and Kavukcuoglu, Koray},
  journal={arXiv preprint arXiv:1609.03499},
  year={2016}
}

@article{kong2020hifi,
  title={Hifi-gan: Generative adversarial networks for efficient and high fidelity speech synthesis},
  author={Kong, Jungil and Kim, Jaehyeon and Bae, Jaekyoung},
  journal={Advances in neural information processing systems},
  volume={33},
  pages={17022--17033},
  year={2020}
}

@inproceedings{shi2021aishell,
  title={AISHELL-3: A Multi-Speaker Mandarin TTS Corpus},
  author={Shi, Yao and Bu, Hui and Xu, Xin and Zhang, Shaoji and Li, Ming},
  booktitle={International Speech Communication Association (Interspeech)},
  year={2021},
  pages={2756--2760}
}

@article{gao2022paraformer,
  title={Paraformer: Fast and accurate parallel transformer for non-autoregressive end-to-end speech recognition},
  author={Gao, Zhifu and Zhang, Shiliang and McLoughlin, Ian and Yan, Zhijie},
  journal={arXiv preprint arXiv:2206.08317},
  year={2022}
}

@article{wang2021vqmivc,
  title={Vqmivc: Vector quantization and mutual information-based unsupervised speech representation disentanglement for one-shot voice conversion},
  author={Wang, Disong and Deng, Liqun and Yeung, Yu Ting and Chen, Xiao and Liu, Xunying and Meng, Helen},
  journal={arXiv preprint arXiv:2106.10132},
  year={2021}
}

@article{Liu2021,
  author = {Liu, S. and Cao, Y. and Wang, D. and others},
  title = {Any-to-many voice conversion with location-relative sequence-to-sequence modeling},
  journal = {IEEE/ACM Transactions on Audio, Speech, and Language Processing},
  year = {2021},
  volume = {29},
  pages = {1717--1728},
}

@article{devlin2018bert,
  title={Bert: Pre-training of deep bidirectional transformers for language understanding},
  author={Devlin, Jacob and Chang, Ming-Wei and Lee, Kenton and Toutanova, Kristina},
  journal={arXiv preprint arXiv:1810.04805},
  year={2018}
}

@article{du2024cosyvoice,
  title={Cosyvoice: A scalable multilingual zero-shot text-to-speech synthesizer based on supervised semantic tokens},
  author={Du, Zhihao and Chen, Qian and Zhang, Shiliang and Hu, Kai and Lu, Heng and Yang, Yexin and Hu, Hangrui and Zheng, Siqi and Gu, Yue and Ma, Ziyang and others},
  journal={arXiv preprint arXiv:2407.05407},
  year={2024}
}

@article{kong2023vits2,
  title={VITS2: Improving quality and efficiency of single-stage text-to-speech with adversarial learning and architecture design},
  author={Kong, Jungil and Park, Jihoon and Kim, Beomjeong and Kim, Jeongmin and Kong, Dohee and Kim, Sangjin},
  journal={International Speech Communication Association (Interspeech)},
  year={2023}
}

@inproceedings{xiao2020improving,
  title={Improving prosody with linguistic and bert derived features in multi-speaker based mandarin chinese neural tts},
  author={Xiao, Yujia and He, Lei and Ming, Huaiping and Soong, Frank K},
  booktitle={ICASSP 2020-2020 IEEE International Conference on Acoustics, Speech and Signal Processing (ICASSP)},
  pages={6704--6708},
  year={2020},
  organization={IEEE}
}

@article{zhang2021extracting,
  title={Extracting and predicting word-level style variations for speech synthesis},
  author={Zhang, Ya-Jie and Ling, Zhen-Hua},
  journal={IEEE/ACM Transactions on Audio, Speech, and Language Processing},
  volume={29},
  pages={1582--1593},
  year={2021},
  publisher={IEEE}
}

@article{cui2021pre,
  title={Pre-training with whole word masking for chinese bert},
  author={Cui, Yiming and Che, Wanxiang and Liu, Ting and Qin, Bing and Yang, Ziqing},
  journal={IEEE/ACM Transactions on Audio, Speech, and Language Processing},
  volume={29},
  pages={3504--3514},
  year={2021},
  publisher={IEEE}
}

@article{wang2017tacotron,
  title={Tacotron: Towards End-to-End Speech Synthesis},
  author={Wang, Yuxuan and Skerry-Ryan, RJ and Stanton, Daisy and Wu, Yonghui and Weiss, Ron J and Jaitly, Navdeep and Yang, Zongheng and Xiao, Ying and Chen, Zhifeng and Bengio, Samy and others},
  journal={Interspeech 2017},
  pages={4006},
  year={2017},
  publisher={ISCA}
}

@article{ren2019fastspeech,
  title={Fastspeech: Fast, robust and controllable text to speech},
  author={Ren, Yi and Ruan, Yangjun and Tan, Xu and Qin, Tao and Zhao, Sheng and Zhao, Zhou and Liu, Tie-Yan},
  journal={Advances in neural information processing systems},
  volume={32},
  year={2019}
}

@article{kim2020glow,
  title={Glow-tts: A generative flow for text-to-speech via monotonic alignment search},
  author={Kim, Jaehyeon and Kim, Sungwon and Kong, Jungil and Yoon, Sungroh},
  journal={Advances in Neural Information Processing Systems},
  volume={33},
  pages={8067--8077},
  year={2020}
}

@article{borsos2023audiolm,
  title={Audiolm: a language modeling approach to audio generation},
  author={Borsos, Zal{\'a}n and Marinier, Rapha{\"e}l and Vincent, Damien and Kharitonov, Eugene and Pietquin, Olivier and Sharifi, Matt and Roblek, Dominik and Teboul, Olivier and Grangier, David and Tagliasacchi, Marco and others},
  journal={IEEE/ACM transactions on audio, speech, and language processing},
  volume={31},
  pages={2523--2533},
  year={2023},
  publisher={IEEE}
}

@article{wang2023neural,
  title={Neural codec language models are zero-shot text to speech synthesizers},
  author={Wang, Chengyi and Chen, Sanyuan and Wu, Yu and Zhang, Ziqiang and Zhou, Long and Liu, Shujie and Chen, Zhuo and Liu, Yanqing and Wang, Huaming and Li, Jinyu and others},
  journal={arXiv preprint arXiv:2301.02111},
  year={2023}
}

@article{liu2024zero,
  title={Zero-shot voice conversion with diffusion transformers},
  author={Liu, Songting},
  journal={arXiv preprint arXiv:2411.09943},
  year={2024}
}
\end{document}